\documentclass[a4paper,aps,prl,twocolumn,groupedaddress]{revtex4}
\usepackage{graphicx}
\usepackage{bm}

\begin{document}
\pacs{73.63.Kv, 72.25.-b}
\title{Pauli spin-blockade in an InAs nanowire double quantum dot}
\author{A.~Pfund, I.~Shorubalko, R.~Leturcq and K.~Ensslin}

\address{Solid State Physics Laboratory, ETH Z\"urich, 8093 Z\"urich, Switzerland\\
E-mail: pfund@phys.ethz.ch }

\begin{abstract}
We measure transport at finite bias through a double quantum dot formed by top-gates in an InAs nanowire. Pauli spin-bockade is confirmed with several electrons in the dot. This is expected due to the small exchange interactions in InAs and the large singlet-triplet splitting, which can be measured and tuned by a gate voltage.
\end{abstract}

\maketitle
Spin states in coupled quantum dots are extensively discussed as possible realizations for quantum bits in solid state based quantum computers \cite{PhysRevA.57.120}. Spin correlation leads to distinguishable charge distributions for the different spin states and provides means for electric read-out and manipulation of the states. Pauli spin-blockade (SB) \cite{Ono_rectification} has been employed for this purpose to manipulate singlet (S) and triplet (T) states \cite{PettaScience2005,Koppens:2006fk} and to study spin relaxation mechanisms in GaAs-based double quantum dots \cite{Koppens:2005qy,johnsonPulse}.

We observe SB in a double quantum dot (DQD) defined by top-gates along an InAs nanowire (NW). Pauli spin-blockade for several electrons is identified not only by its rectification effect \cite{Ono_rectification}, but also by the characteristic current variations on small magnetic field scales \cite{Koppens:2005qy}. 

Our device is fabricated using InAs nanowires grown by metal organic vapor phase epitaxy from a catalytic Au nanoparticle \cite{hiruma:447}. The NW is contacted on a Si substrate with an insulating surface oxide layer and metallic top-gates are lithographically defined as shown in Fig.\,1a) \cite{pfund:252106}. Transport measurements are performed in a dilution refrigerator at an electronic temperature of $\sim100\,$mK and with a magnetic field aligned perpendicular to the nanowire axis.
\begin{figure}
\begin{center}\leavevmode
\includegraphics[width=0.4\textwidth]{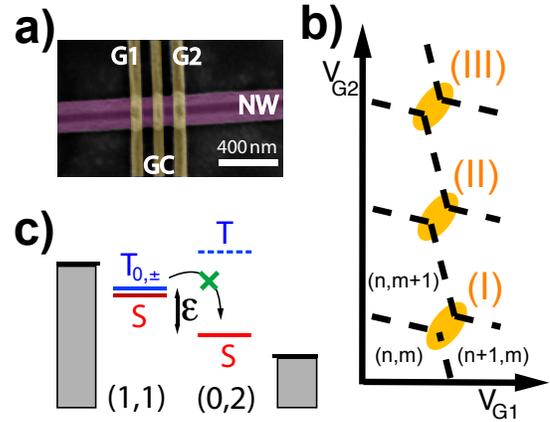}
\caption{a) Scanning electron microscope image of the device. Nanowire (NW) deposited on Si with 300$\,$nm Si-oxide top layer. Three Cr/Au top-gates are defined by electron beam lithography and standard lift-off. b) Sketch of the charge stability diagram as a function of gates $G1,G2$ which tune energy levels in dot $1,2$. Sequential transport is only allowed around triple points (I)-(III). c) Energy diagram for transport through a two-electron double dot at finite bias (tunnel coupling neglected). For small detuning $\varepsilon$ between $(1,1)$ and $(0,2)$ states, Pauli spin-blockade suppresses sequential transport once the second electrons enters in a $(1,1)$-triplet.}
\end{center}
\end{figure}

A DQD is electrically defined by gates $G1,G2$ and $GC$. The inter-dot coupling is adjusted with $GC$. Gates $G1,G2$ tune the electrochemical potentials in dot 1,2 respectively. The electronic occupation $(n,m)$ is fixed in characteristic honeycombs in the $V_{G1}$-$V_{G2}$-plane \cite{vanderWiel01}. Strong current due to sequential transport is observed only around ``triple points'', where the dot levels are degenerate with electrochemical potentials of source or drain. Fig.\,2 shows the current for positive and negative bias around the 3 neighboring pairs of triple points named (I)-(III) as highlighted in Fig.\,1b). Transport is in principle allowed in triangular regions where the corresponding dot levels are within the bias window \cite{vanderWiel01}.
\begin{figure}
\begin{center}\leavevmode
\includegraphics[width=0.4\textwidth]{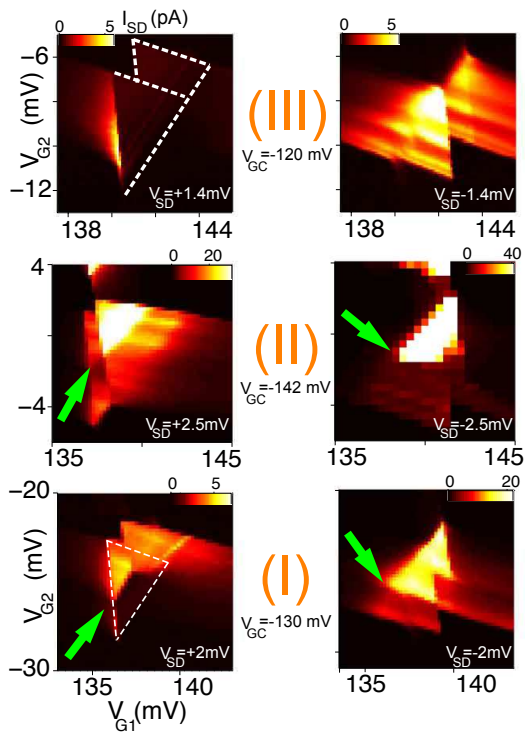}
\caption{Double dot current around the 3 degeneracy points (I)-(III) as outlined in Fig.\,1b) for positive (left column) and negative bias (right column). The coupling has been adjusted using $GC$ in the 3 cases, which explains the variation in the voltages for $G1,G2$.
Spin blockade is observed for (I) and (III) where the electronic occupation of dot 2 differs by two. The excited state is consistently observed in both bias directions for (I) and (II) (arrows), while it is outside of the bias window in (III).}
\end{center}
\end{figure}

Around (I) and (III), the current is suppressed in the base region of the triangles for positive bias direction. For reverse bias, the current level is always high. The two configurations differ by 2 electrons added to dot 2. For the triple points (II), significant current is observed in both bias directions (note the different current scales).
This observation can be explained by Pauli spin-blockade. In Fig.\,1c) the principle is explained for a  DQD containing one electron in dot 2: Because tunneling preserves spin, the sequential transport cycle (0,1)$\rightarrow$(1,1)$\rightarrow$(0,2)$\rightarrow$(0,1) is blocked, once the second electron enters in a triplet  \cite{Ono_rectification}. If the detuning $\varepsilon$ exceeds the ST-splitting of the (0,2)-states, transport is allowed again. This gives rise to a sharp step in current. For reverse bias, transport is not excluded by spin conservation.

In our case, SB relies on the existence of an excited triplet state in dot 2 with a level splitting which is larger than other splittings of the involved states. In case (I), this ST-splitting is consistently seen for both bias directions (see arrows). Adding another electron to dot 2 leads to the situation around (II). No SB is expected in this configuration and current is allowed for both bias directions. Once the excited state enters the bias window, the current increases due to the additional transport channel \cite{vanderWiel01}. The corresponding current step is again visible for positive and negative bias. In (III), the bias window is smaller than the ST-splitting and no step in current occurs (the current onset is clearly observed for $V_{SD}>2mV$, not shown here).

In all three cases (I)-(III), strong cotunneling through dot 1 gives rise to stripes of background current roughly independent on $V_{G1}$. This process obscures SB in the upper right part of the triangles for (I). As discussed in the following, SB can be more clearly identified by its characteristic sensitive magnetic field dependence.

In Fig.\,3a), the triangles around (I) are shown for weaker inter-dot coupling. Here, spin relaxation processes induced by, for example, spin-orbit or hyperfine interaction partly destroy SB due to mixing of $(1,1)$ singlets and triplets among each other \cite{Koppens:2005qy,johnsonPulse}.
Fig.\,3b) shows the current as a function of detuning and magnetic field for the three regions (I)-(III). The gates are varied along detuning lines inside the triangles as indicated in Fig.\,3a). Gate voltages are translated into level detuning $\varepsilon$ using leverarms which can be obtained by comparing the dimensions of the triangles with the applied bias voltage \cite{vanderWiel01}. Only in the cases (I) and (III) we observe a variation of the current on a small field scale. This behavior is consistent with previous observations in spin-blockaded DQDs \cite{Koppens:2005qy,pfund-2007-HFSO}: a magnetic field splits the $(1,1)$-triplets and the mixing among the $(1,1)$-states is reduced.
A striking sensitive B-dependence is also observed for strong inter-dot coupling, where relaxation in SB leads to dynamic polarization of nuclear spins \cite{pfund-2007-DNSP}.
Consistently, we find no magnetic field dependence in case (II), where SB is not expected to occur.
\begin{figure}
\begin{center}\leavevmode
\includegraphics[width=0.4\textwidth]{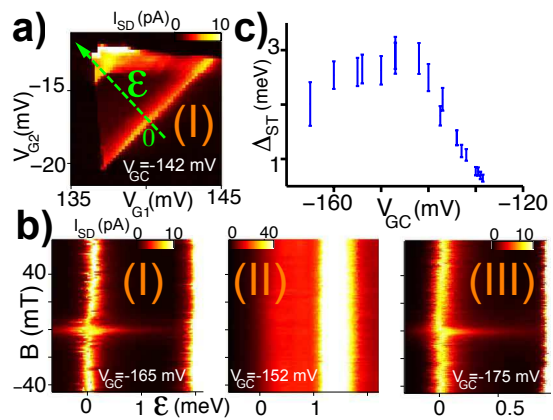}
\caption{a) For weaker coupling, spin-blockade is lifted by spin relaxation. $V_{SD}=4\,$mV. b) Relaxation current as a function of magnetic field and detuning $\varepsilon$ of $(1,1)$ and $(0,2)$ levels. A characteristic variation of the current on these small field scales is observed only for configurations showing spin-blockade (I) and (III). c) Splitting between S(0,2) and T(0,2) as determined from the spin-blockaded triangles for different $V_{GC}$. Bias voltage $3.5\,$mV,$\,2.5\,$mV,$\,1.5\,$mV for (I),(II),(III) respectively.}
\end{center}
\end{figure}

In our measurement, most likely more than one electron occupies the DQD around the triple points (I). 
For DQDs containing 2 electrons, SB is expected because the 2 electron ground state is a singlet. For more electrons, exchange interactions may become comparable to the level spacing and create a ground state with higher spin \cite{PhysRevB.66.195314}. Nevertheless, SB was observed in GaAs DQDs containing up to around 10 electrons \cite{johnson:165308,liu:161305}.
In InAs the small effective mass leads to large level splittings and small exchange interactions and the formation of spin-pairs is therefore likely also for larger electron numbers. This could explain the occurrence of SB for configurations with more than 2 electrons.

In the spin-blockade configuration, we can identify the singlet and triplets in dot 2 and study the dependence on external parameters. Fig.\,3c) shows the \mbox{$V_{GC}$-dependence} of the singlet-triplet splitting $\Delta_{ST}$, which is extracted from the spin-blockaded triangles for different $V_{GC}$.
Changing the gate voltages alters the electrostatic confinement of the dots and therefore $\Delta_{ST}$ \cite{PhysRevB.66.035320}. A more negative $V_{GC}$ in principle implies stronger confinement for the separate dots (assuming rigid outer barriers). This explains qualitatively the increase of $\Delta_{ST}$ with decreasing $V_{GC}$ starting from about $-130\,$mV. The dependence is consistent with results for lateral quantum dots \cite{meunier:126601}, where the singlet-triplet splitting changes linearly in gate voltage. For $V_{GC}<-145\,$mV we observe a change in the behavior, which could be related to a change in confinement.\\
As expected for InAs, the measured level spacing in the meV-range is large compared to other energy scales of the dot \cite{pfund-2007-HFSO}.

In summary, we detected Pauli spin-blockade in an InAs nanowire DQD containing several electrons. Evidence is provided by the rectification behavior, the characteristic magnetic field dependence and the observed large level splitting.

We thank M.~Borgstr\"om and E.~Gini for advice in nanowire growth. We acknowledge financial support from ETH Zurich and I.S. thanks the European Commission for a Marie-Curie fellowship.

\end{document}